\documentclass{aa} % 
\usepackage{graphicx}
\usepackage{txfonts}
\usepackage{color}

\newcommand{\markmeold}[1]{#1}
\newcommand{\markme}[1]{#1}
\newcommand{\srem}[1]{}
\usepackage{psfrag,pstricks,epsfig,tabularx}
\usepackage{ulem}
\begin{document} 
\title{Observations of high and low Fe charge states in individual solar wind streams with coronal-hole origin}
\titlerunning{Fe charge states in \markmeold{coronal-hole wind}}
\author{Verena Heidrich-Meisner \inst{1}, Thies Peleikis \inst{1}, Martin Kruse \inst{1}, Lars Berger \inst{1}, and Robert Wimmer-Schweingruber \inst{1}}

\authorrunning{V. Heidrich-Meisner, T. Peleikis, M. Kruse, L. Berger, and R. Wimmer-Schweingruber}
%\authorrunning{V. H.-M., T. P., M. K., L. B.,  R. W.-S.}

\institute{Christian Albrechts University at Kiel, Germany,
  \email{heidrich@physik.uni-kiel.de}
}

\date{}

\abstract{The solar wind originating from coronal holes is comparatively well-understood and is characterized by lower densities and average charge states compared to the so-called slow solar wind. Except for wave perturbations, the {average properties of the coronal-hole solar wind are \markmeold{passably} constant.}}{{In this case study}, we {focus on} observations of the Solar Wind Ion
Composition Spectrometer (SWICS) on the Advanced Composition Explorer (ACE) of {individual streams of coronal-hole solar wind} that {illustrate that although the O and C charge states are low in coronal-hole wind, the Fe charge distribution is more variable. In particular, \markmeold{we illustrate that} the Fe charge states in coronal-hole solar wind are frequently as high \markmeold{as} in slow solar wind.}}{\markmeold{We} selected {individual coronal-hole solar wind} streams based on their collisional age as well as their respective O and C charge states and analyzed their Fe charge-state distributions. {Additionally, with a combination of simple ballistic back-mapping and the potential field source surface model, transitions between streams with high and low Fe charge states were mapped back to the photosphere.} \markmeold{The relative frequency of high and low Fe charge-state streams is compared for the years 2004 and 2006.} }{\markmeold{We found several otherwise typical coronal-hole streams that include Fe charge states either as high as or lower than in slow solar wind.  Eight such transitions in 2006 were mapped back to equatorial coronal holes that were either isolated or connected to the northern coronal-hole. Attempts to identify coronal structures associated with the transitions were so far inconclusive.}}{} 
% 5 {} token are mandatory
 

   \maketitle
%
%________________________________________________________________

\section{Introduction}
The steady solar wind is typically divided into \markmeold{two dominant} types, fast and slow solar wind. {However, the differences in their properties are better ordered by elemental and charge-state abundances rather than by solar wind speed.} Here, we focus on fast solar wind. It has been uniquely identified as originating from coronal holes and the release mechanism is well understood (i.e. \citet{Tu2005}). Therefore, in the following, we use the term {coronal-hole wind} instead of fast solar wind. Aside from fluctuations caused by waves (mainly Alfv{\'e}nic waves), its plasma and compositional properties are \srem{tolerably} constant \markmeold{(e.g. \citet[]{steiger2000composition})}. 
Both the elemental and charge-state compositions of the solar wind reflect the conditions in the respective solar source regions. In particular, the charge-state distribution for each solar wind ion species is (almost completely) determined in the corona.
For each ion pair the recombination and ionization rates are temperature dependent and the hot corona allows high ionization states to occur.  For a pair of adjacent ionization states $i\leftrightarrow i+1$, this can be expressed by a temperature dependent {charge} modification time scale $\tau_{\text{mod},i}(T) = \frac{1}{n_e(C_i+R_{i+1})}$, where $T$ denotes the electron temperature in K, $n_e$ the electron density\markmeold{,} $C_i$ the ionization rate of the $i$-th ionization state, and $R_{i+1}$ the recombination rate of the $(i+1)$th to the $i$th ionization state. But the ionization state \markmeold{is} not only temperature dependent. A sufficiently high electron density is required to allow recombination. Thus, a simple model to explain the observed solar wind speeds and charge states assumes that the charge state can change along the solar-wind flux tube until the expansion time scale (which depends on the electron density profile in the corona) is of the same order as the charge modification time scale of an ion pair. Beyond this point the charge-state distribution remains ``frozen-in'' as the solar wind propagates further outwards.

The {coronal-hole wind} is known for comparatively low O and C charge states and corresponding freeze-in temperatures. They are particularly low compared to those of the slow solar wind. The \markmeold{O} charge-state distribution \markmeold{can be \markmeold{considered} as a} tracer for the solar wind type. \markmeold{The} ratio {$n_{O^{7+}}/n_{O^{6+}}$} of the {densities} of $O^{7+}$ to $O^{6+}$ (denoted with $n_{O^{7+}}$ and $n_{O^{6+}}$, respectively) has been frequently used in solar wind categorization schemes \markmeold{(e.g. \citet[]{zurbuchen2002solar,zhao2009global})} to differentiate between fast (coronal hole) and slow solar wind.
In accordance with the cool O and C signatures in {coronal-hole wind}, the charge-state distributions of other ions, for example Fe, could also be expected to be cooler in the {coronal-hole wind} than in the slow solar wind. However, that is clearly not necessarily the case.%(see e.g. \citet[]{steiger2000composition,galvin2009solar,richardson2014identification,lepri2013solar})
 While the Ulysses observations in \citet[]{steiger2000composition}, \citet[]{richardson2014identification} \markmeold{and} \citet[]{zhao2014polar} (as well as the ACE observations in \cite[]{zhao2014polar}) \markmeold{show} on average higher Fe charge states in the fast solar wind than in the slow solar wind (see for example Plate~5 in \citet[]{steiger2000composition}), the STEREO results \markmeold{(\citet[]{galvin2009solar}) indicate} lower Fe charge states \markmeold{in coronal-hole wind}.  For example, for the solar-wind speed bin $650km/s-700km/s$ the average Fe charge state in coronal-hole wind observed with PLASTIC on STEREO A from 2007-2009 is given as $9.25$, which is lower than for all slow solar wind bins considered in that article. For all other high speed bins, the average Fe charge state is even lower. 
\citet[]{zhao2014polar} compared long-term properties of coronal-hole wind at different solar minima and identified two subcategories, coronal-hole wind originating from polar coronal holes and coronal-hole wind originating from the equatorial region. Lower Fe charge states were observed in equatorial coronal-hole wind than in polar coronal holes and additionally lower charge states were found in the second solar minimum.
A gradual charge-state decrease for O, C, and Fe from solar maximum to the following solar minimum in solar cycle 23 has been discussed in \citet[]{lepri2013solar} and \citet[]{zhao2010comparison} underlines differences in \markmeold{the composition of the slow solar wind between \markmeold{the} two consecutive solar minima \markmeold{in solar cycles 22 and 23}.}

Instead of focusing on the statistics of charge-state parameters gathered over long time periods, \markmeold{we discuss} several case studies of individual streams \markmeold{within the ACE/SWICS data} that show regions of high and low Fe charge states with a clear transition between these regions. We then map \markmeold{these} streams back to their coronal sources and assess whether there are any coronal structures that may be associated with these transitions.

\section{\markmeold{Data analysis and event selection}}\label{sec:data}
The SWICS instrument on ACE \markmeold{\citep{gloeckler-etal-1998}} combines a time-of-flight mass spectrometer and energy-per-charge analyzer {with} an energy measurement. A detailed and extensive description of the data analysis procedure applied{ to the pulse height amplitude (PHA) data} is given in the PhD thesis \citet[]{berger2008velocity} { and has been applied in, e.g. \citet{berger2011systematic}.}

\begin{figure}[tb!]\centering
\includegraphics[width=0.8\columnwidth]{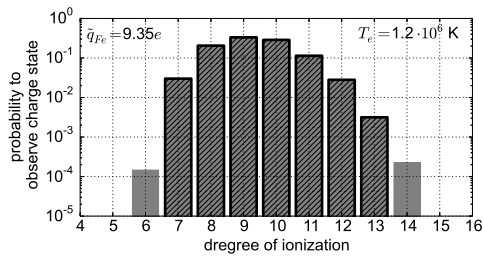}
\includegraphics[width=0.8\columnwidth]{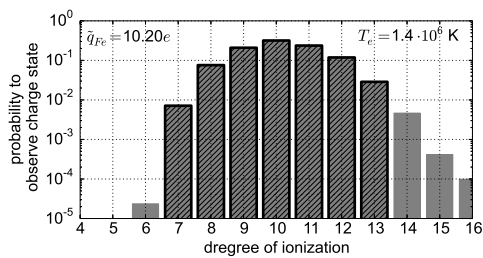}
\caption{\label{fig:chiantiCS} Top panel: Fe charge-state distribution at $T=1.2 \times 10^6 K$ taken from the CHIANTI database. Bottom panel: Fe charge-state distribution at $T=1.4 \times 10^6 K$ from the CHIANTI. The black borders indicate the Fe charge states considered in this work. In both panels, the mean Fe charge-state $\tilde{q}_{Fe}$ is given as inset on the left.}
\end{figure}

The proton density $n_p$ (from the Solar Wind Electron, Proton and Alpha Monitor (SWEPAM) on ACE \markmeold{, \citet{mccomas1998solar}}), and the densities of
$O^{6+}$, $O^{7+}$, $C^{5+}$, and $C^{6+}$ are used for
the characterization of solar-wind plasma as {coronal-hole wind}.  We selected
{coronal-hole-wind} streams based on {four-hour
  resolution} data from SWICS and SWEPAM and the following three criteria:
\markmeold{(1)} \markmeold{Low} O and C charge-state \markmeold{ratios
  ($n_{O^{7+}}/n_{O^{6+}}$ and $n_{C^{6+}}/n_{C^{5+}}$, respectively)}
are considered as the decisive property for identifying
{coronal-hole wind}. \citet{zhao2009global} proposed an
upper limit of {$n_{O^{7+}}/n_{O^{6+}}$}$<0.145$ for
{coronal-hole wind}. To avoid potential contamination with
inter-stream solar wind, we apply
{$n_{O^{7+}}/n_{O^{6+}}$}$<0.1$ which was also used in
\citet{zurbuchen2002solar}. For C, we adopted an upper threshold of
$n_{C^{6+}}/n_{C^{5+}}<1$. \markmeold{(2)} \markmeold{Based on the observations in \citet{kasper2008hot}, we additionally} require a low
collisional age $a_{\text{col}}=\frac{r}{v_{p}  \tau_{\markmeold{\text{col}}}}<0.1$, with $r$ as the distance from the
Sun to ACE, $v_p$ the solar-wind proton speed, and $\tau_{\text{col}}$
the time scale for $\alpha$ to proton energy exchange due to
small-angle Coulomb scattering, \markme{where} $\tau_{\text{col}} \sim n_p T_p^{-3/2}$. The collisional age is defined by the
ratio of expansion and collisional time scales. Although the
\markmeold{proton-proton} collisional age is not representative for the
collisional properties of the entire solar-wind plasma, we consider
the \markmeold{proton-proton} collisional age as a representative marker for the solar-wind
stream type. \markmeold{(3) Only streams that remained within the
  respective same categorization regimes of the average O and C charge states and collisional age} for at least half a day were considered.
Fluctuations \markmeold{in $a_{\text{col}}$, $n_{O^{7+}}/n_{O^{6+}}$,
  $n_{C^{6+}}/n_{C^{5+}}$, or the average Fe charge-state }on smaller
time scales than the four hours were permitted as long as the average
value (averaged over four hours) remained in the respective range.

\markmeold{As a comparison baseline we also require pure slow
    solar wind.  In this context, pure slow solar wind is
  characterized by high O charge states $n_{O^{7+}}/n_{O^{6+}}>0.1$,
  high C charge states $n_{C^{6+}}/n_{C^{5+}}>1$, and high collisional
  age $a_{\text{col}}>0.4$. This characterization of slow solar wind is not directly
  complementary to the criteria for identifying coronal-hole wind as described above in order to
  reduce the contamination of each wind type by transition regions
  that exhibit a mixture of properties of slow and coronal-hole wind.}

%Figure~\ref{categorizeexample} provides an example of the categorization scheme. Day of year (DoY) \verena{add this} The black lines in all panels represent 4-hour resolution data, the underlying gray lines represent 12-minute resolution data. Only the parameters used in the categorization are shown here, the ratio $n_{O^{7+}}/n_{O^{6+}}$ in the first panel, the collisional age $a_{\text{col}}$ in the second panel.

In this study, we are interested in the Fe charge-state distribution in {coronal-hole wind}. The densities of the Fe charge states are provided by ACE/SWICS. In particular, since the most abundant Fe charge states are well isolated from all other ions in the SWICS $m$-$m/q$ diagram, we focus on these, namely $Fe^{7+}$, $Fe^{8+}$, $Fe^{9+}$, $Fe^{10+}$, $Fe^{11+}$, $Fe^{12+}$, and $Fe^{13+}$. In the following the average Fe charge-state $\tilde{q}_{Fe}$ is defined as $\tilde{q}_{Fe} = \sum_{c=7}^{13} c n_{Fe^{c+}}/ \sum_{c=7}^{13} n_{Fe^{c+}}$.

{Figure~\ref{fig:chiantiCS} shows the Fe charge-state distribution for two electron temperatures $T=1.2 \times 10^6 K$ (top) and $T=1.4 \times 10^6 K$ (bottom) as provided by the atomic database CHIANTI \citep[]{dere1997chianti,landi2013chianti}. \markmeold{The temperatures are chosen from the typical range of observed electron temperatures in the corona (see e.g. \cite{ko1997empirical,wilhelm2012sumer}).} The charge states considered here ($Fe^{7+}$, $Fe^{8+}$, $Fe^{9+}$, $Fe^{10+}$, $Fe^{11+}$, $Fe^{12+}$, and $Fe^{13+}$) are highlighted by hatched bars and black borders around the respective bars and are at the relevant temperatures the most prominent charge states.}

{
Based on the CHIANTI data displayed in Figure~\ref{fig:chiantiCS} and under the assumptions that the freeze-in temperature $T_{f, Fe}$ for all Fe ions is the same and that $T_{f,Fe} \sim1.2 \times 10^6 K$, Figure~\ref{fig:chiantiCS} also illustrates that $q_{Fe}=9$ would be the most likely charge state, with a mean charge state of $9.35$. For $T=1.4 \times 10^6 K$ a mean charge state of $10.2$ would be expected. However, since the assumption that all Fe ions freeze-in at the same temperature is not accurate, this provides only a rough guideline. \markmeold{Furthermore} CHIANTI makes the assumption of a Maxwellian distribution for the electron velocity distribution function which is known not to be the appropriate choice for the solar corona and the solar wind (see for example, \citet[]{marsch2006kinetic}). Instead our notion of high or low average Fe charge state is based on a comparison of Fe charge states in coronal-hole wind to those observed in \markmeold{selected samples of} slow solar wind.}

%\clearpage

For 2004 the median of the average Fe charge-state of pure slow solar wind was $\tilde{q}_{Fe, slow}=9.87$. The $1\sigma$-level is bounded below by $\tilde{q}_{Fe, slow} - \sigma= 9.54$. These values are used to define our notion of high and low Fe charge states. We consider the average Fe charge state $\tilde{q}_{Fe, CH}$ of a coronal-hole-wind stream as low and the stream as Fe-cool if $\tilde{q}_{Fe, CH}< \tilde{q}_{Fe, slow}-\sigma$. Analogously, we consider a coronal-hole-wind stream to be Fe-hot if its average Fe charge-state is within one $\sigma$ of the average Fe charge state of pure slow solar wind for that year or higher: $\tilde{q}_{Fe, CH} > \tilde{q}_{Fe, slow} - \sigma$. \markmeold{Thus} for 2004, the threshold value is $\tilde{q}_{Fe, slow}-\sigma=9.54$ and for 2006 $\tilde{q}_{Fe, slow}-\sigma=9.71$.

{To ensure that the selected coronal-hole-wind streams are not contaminated with interplanetary coronal mass ejections (ICMEs), w}e cross-referenced the \citet{jian2006properties,jian2011comparing} and \citet{richardson2010near} ICME lists and the Large Angle and Spectrometric Coronagraph (LASCO) CME
list \markmeold{and} excluded all time periods with ICMEs from our times of interest\markmeold{.}
%
%{, i.e. the time periods that are identified as coronal-hole \markmeold{or pure slow solar wind} according to the criteria defined above.
%
For four days after each halo CME in the LASCO list \markmeold{\citep{yashiro2004catalog,gopalswamy2009soho}} that does not have a counterpart in the ICME lists, we verified that no ICME signatures were contained in the coronal-hole-wind stream candidates. Also, to reduce the effect of inter-stream regions, time periods with enhanced proton density and magnetic field strength which indicate stream interaction-regions \citep{jian2006properties,jian2011comparing} were excluded as well. \markmeold{With this method, 4660 12-minute observations were selected as coronal-hole wind in 2004, accumulating to 38.8 days of combined coronal-hole wind, and 8346 observations corresponding to 69.5 days of combined coronal-hole wind in 2006. For the sake of an unbiased representation, we randomly chose three transitions from two different years between Fe-cool and Fe-hot coronal-hole wind from the available data set\markme{  to be discussed in detail in the following section: day of year (DoY) 2-14 in 2004 and DoY 158-162 and DoY 212-215 in 2006. } \markmeold{Six} additional transitions \markmeold{in} 2006 are mentioned briefly.}

\section{Fe charge states {of individual coronal-hole-wind streams}}

\begin{figure*}[htb!]\centering
  \includegraphics[height=0.8\textheight]{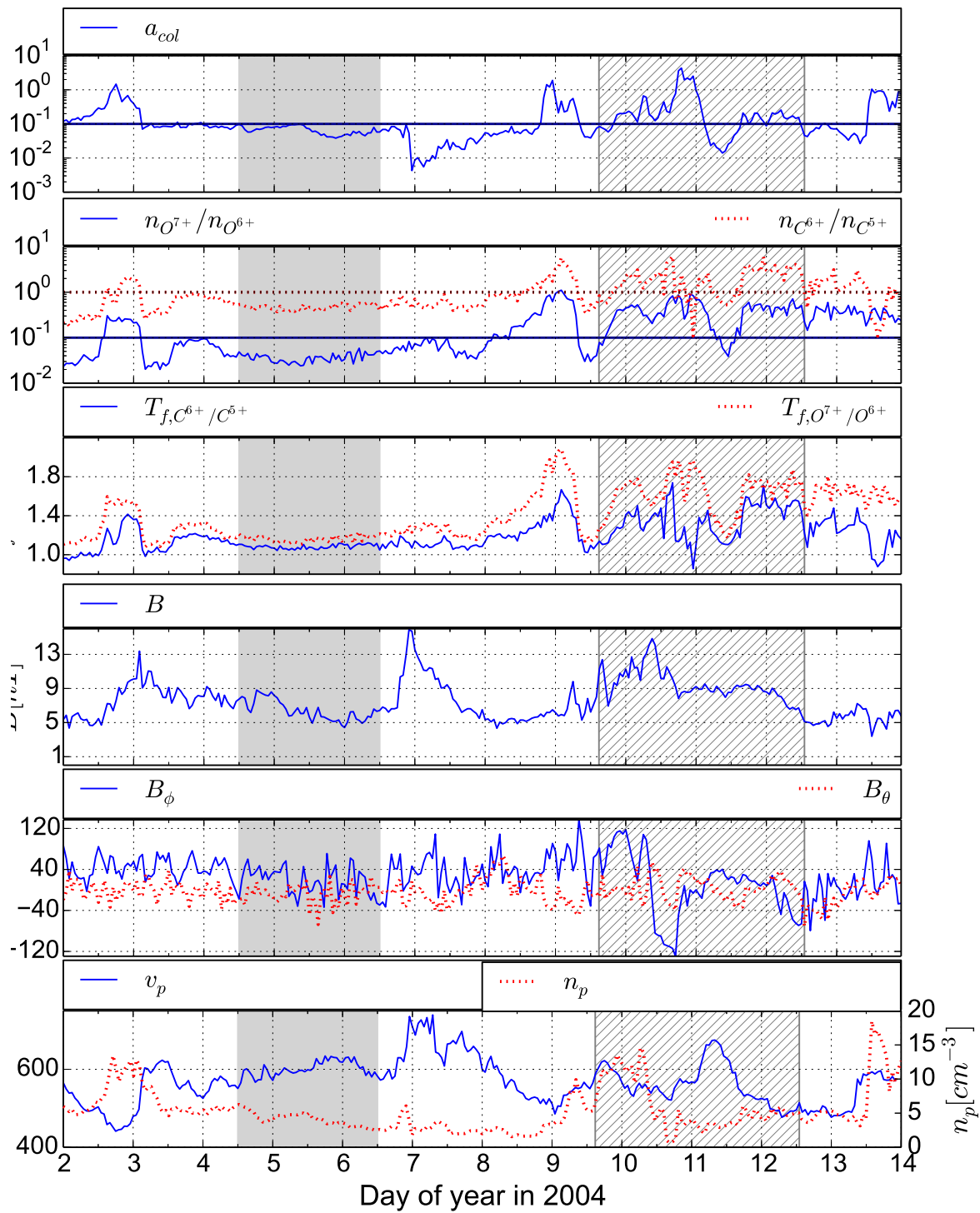}
\caption{\label{fig:coronal} {Solar wind properties for $12$ days in 2004.} The collisional age $a_{\text{col}}$ is displayed in the {first} panel. The {second} panel shows the density ratios \markmeold{$n_{O^{7+}}/n_{O^{6+}}$ and $n_{C^{6+}}/n_{C^{5+}}$}. %The second panel shows the ratio $n_{He^{2+}}/n_{H^+}$.  
 {The horizontal lines in the first two panels indicate the respective selection thresholds for \markmeold{$a_{\text{col}}=0.1$, $n_{C^{6+}}/n_{C^{5+}}=1$, and $n_{O^{7+}}/n_{O^{6+}}=0.1$}.} The {third} panel contains the freeze-in temperatures $T_f$ corresponding to the ion density ratios \markmeold{$n_{O^{7+}}/n_{O^{6+}}$ and $n_{C^{6+}}/n_{C^{5+}}$.} The {fourth} panel gives the \markmeold{magnitude} of the magnetic field $B$ {and the \markmeold{azimuthal ($B_\phi$) and polar ($B_\theta$) angles of the magnetic field} are shown in the fifth panel.} %the FIP ratio $n_{\text{FIP}}$ as the density of the high FIP elements Fe, Mg, and Si divided by the elemental O density compared to the respective photospheric abundance values is used as an indicator for the FIP effect.
 The sixth panel shows the solar-wind proton speed (left y-axis) and proton density (right y-axis) as measured by SWEPAM. All data products are displayed with 1-hour time resolution. {The gray shaded area marks a pure coronal-hole-wind stream, while the hatched area highlights an ICME. }}
\end{figure*}

\begin{figure*}[th!]\centering
  \includegraphics[width=\textwidth, height=0.6\textheight]{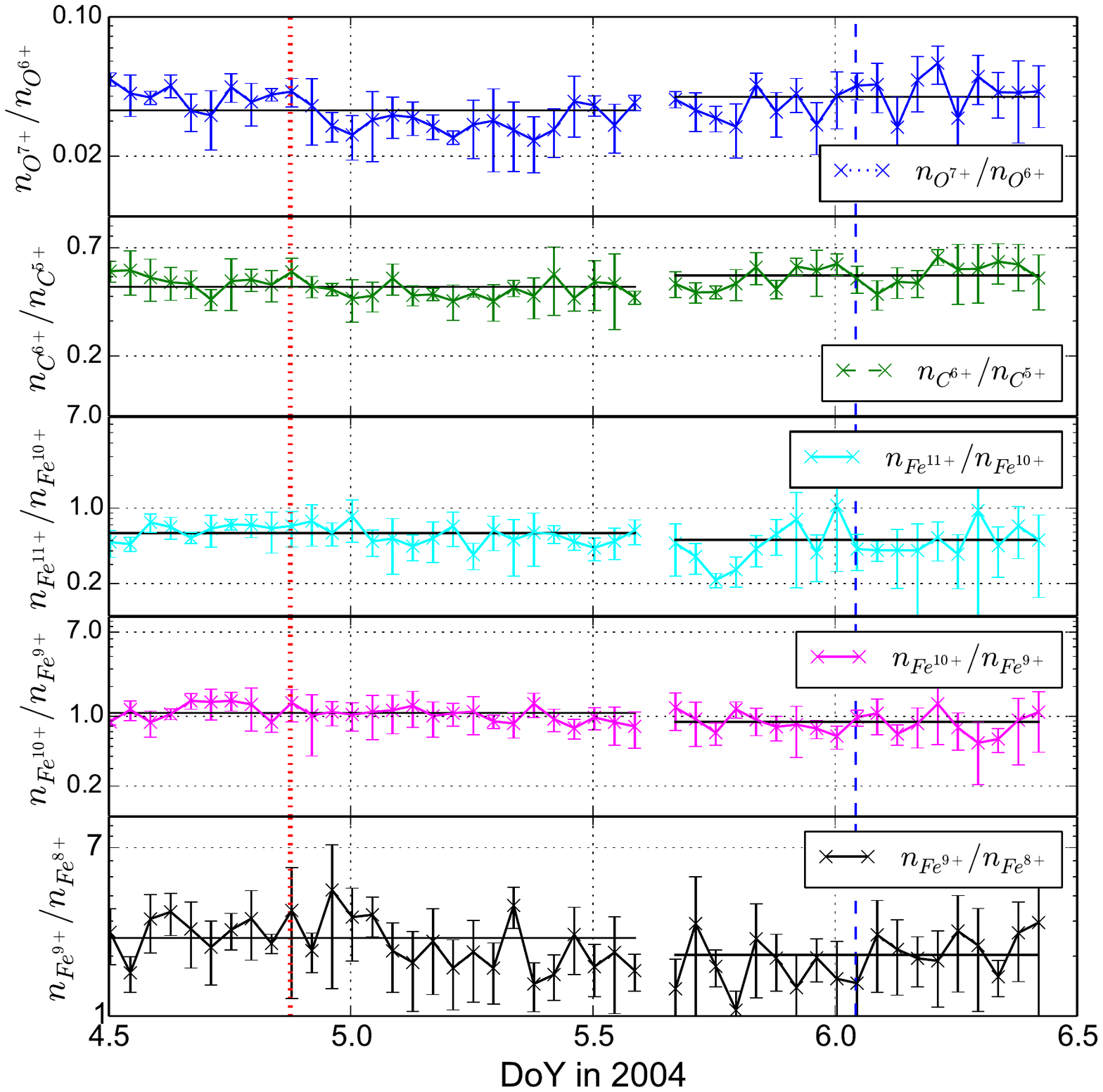}

  \includegraphics[width=\textwidth, height=0.2\textheight]{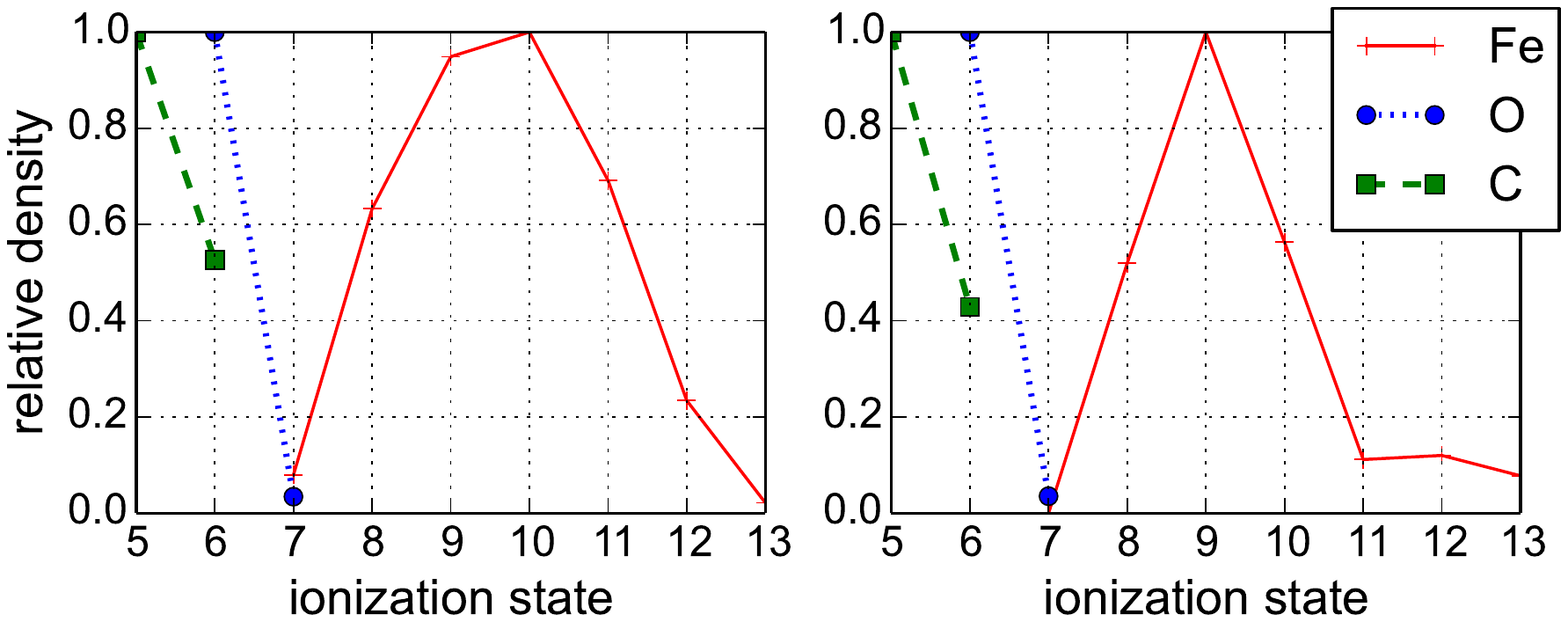}
\caption{\label{fig:examples} \markmeold{Ion density ratios for the highlighted part of the coronal-hole-wind stream from Figure~\ref{fig:coronal}. The five top panels show the density ratios of \markmeold{ion pairs}, namely $n_{O^{7+}}/n_{O^{6+}}$, $n_{C^{6+}}/n_{C^{5+}}$, $n_{Fe^{11+}}/n_{Fe^{10+}}, n_{Fe^{10+}}/n_{Fe^{9+}}, n_{Fe^{9+}}/n_{Fe^{8+}}$. Each curve is divided into two parts. This partition is based on the mean Fe charge-state as illustrated in Figure~\ref{fig:CSTimeseries}. The horizontal lines in each panel give the mean density ratio in the left and right interval, respectively. The vertical lines in the five top panels indicate the points in times for which examples of charge-state distributions normalized to the maximum density for C, O, and Fe \markmeold{are shown} in the bottom panel. The dotted line corresponds to the left bottom panel and the dashed line to the right panel. } (Although $C^{4+}$ is more abundant than $C^{6+}$ it is omitted here because it tends to be contaminated with adjacent O ions in ACE/SWICS.)  Each data point represents a one-hour average and the error bars reflect the error from the counting statistics.}

%\markmeold{{Fe-hot} (left-hand panels) and Fe-cool (right-hand panels) parts of the same coronal-hole-wind stream as in Figure~\ref{fig:coronal}}. Each column shows the ratio of the density $n_X$ of pairs of adjacent ions, namely $n_{O^{7+}}/n_{O^{6+}}$, $n_{C^{6+}}/n_{C^{5+}}$, $n_{Fe^{11+}}/n_{Fe^{10+}}, n_{Fe^{10+}}/n_{Fe^{9+}}, n_{Fe^{9+}}/n_{Fe^{8+}}$\markmeold{,} in the five top panels. \markmeold{}
\end{figure*}
\srem{The data selection method allows {now} to focus on quiet {coronal-hole-wind} streams {and their Fe charge-state distributions.} }

\begin{figure}[t]
  \includegraphics[width=\columnwidth]{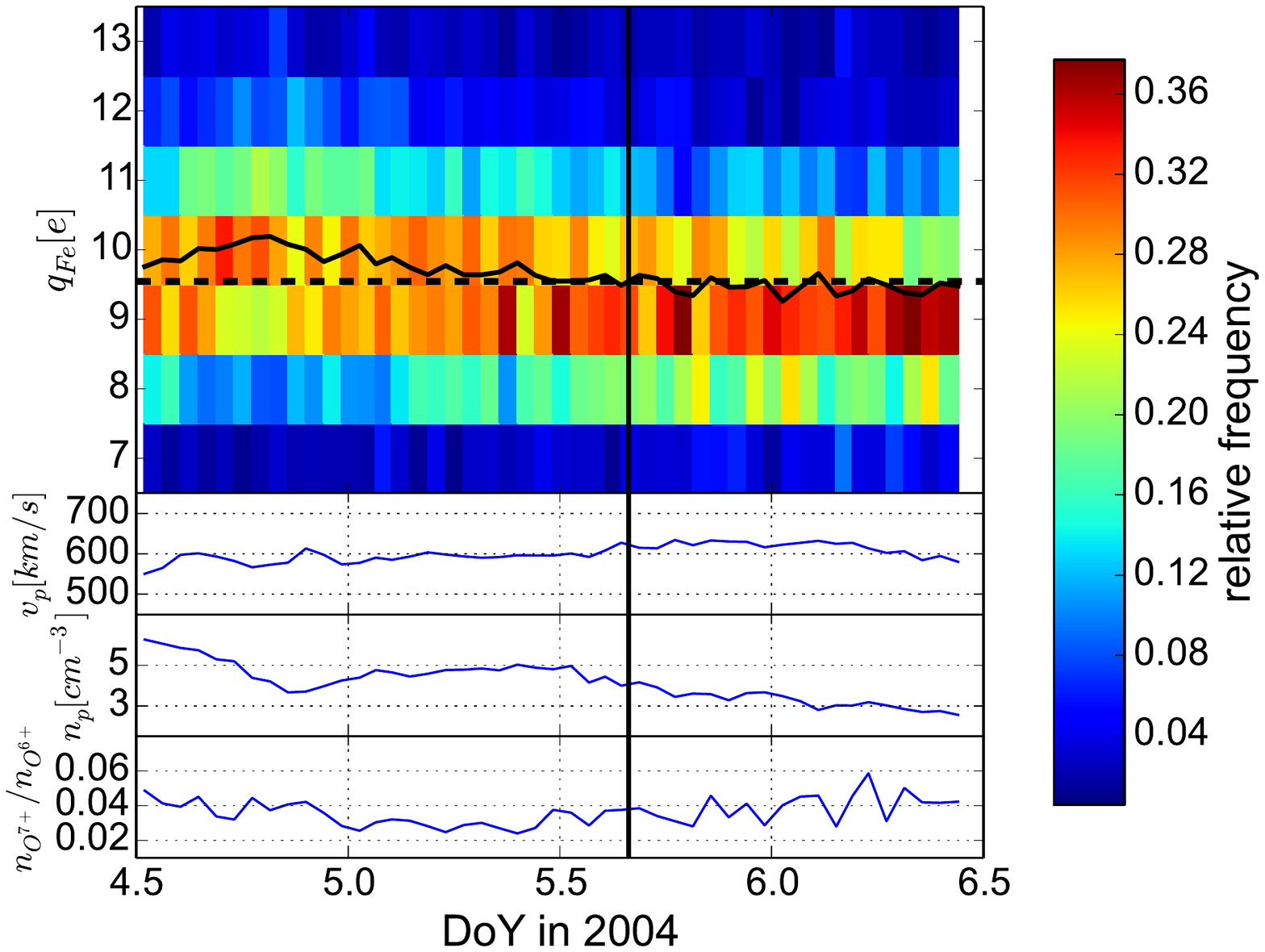}
\caption{\label{fig:CSTimeseries} {Time series of Fe charge-state distributions (first panel) for \markmeold{the highlighted time period from Figure~\ref{fig:coronal}, that is, DoY 4.5-6.5 in 2004,} in one-hour time resolution.} In black, the average charge-state (\markmeold{that is,} $\tilde{q}_{Fe} = \sum_{c=7}^{13} c n_{Fe^{c+}}/ \sum_{c=7}^{13} n_{Fe^{c+}}$) is shown in units of the elementary electric charge $e$. \srem{Each colored block is centred around the charge-state it belongs to.} Each charge-state distribution is normalized to the sum.
{Below}, the solar wind proton speed {(second panel)} and proton density \markmeold{(third panel)} are given as reference. In the bottom panel the ratio {$n_{O^{7+}}/n_{O^{6+}}$} is shown {as well}. \srem{\markmeold{The same period of interest as in Figure~\ref{fig:coronal} is highlighted with gray shading in the three bottom panels. In the first panel the beginning and end of this period is indicated with gray vertical lines.}} {The threshold value between Fe-cool and Fe-hot wind is shown as a horizontal dashed line} and the transition between Fe-hot and {Fe-cool coronal-hole wind} is marked with a vertical black line in all panels.}
\end{figure}

\begin{figure}[htb!]\centering
%\begin{minipage}{0.45\textwidth}
  \includegraphics[width=\columnwidth]{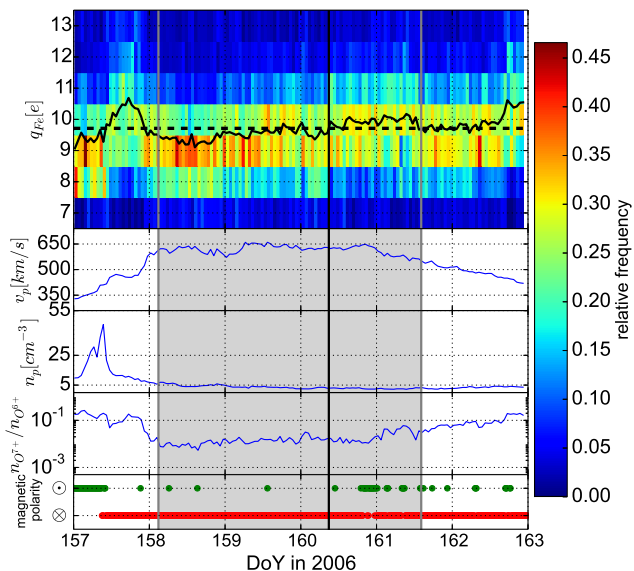}

  %\end{minipage}
%\begin{minipage}{0.45\textwidth}
%  \includegraphics[width=\columnwidth]{backMapped2007_290_295}
  
 %\end{minipage}

%\begin{minipage}{\textwidth}

%  \includegraphics[angle=-90,width=\columnwidth]{map_with_inset}
%\end{minipage}

\caption{\label{fig:coronal-hole-origin} {Time series of the charge-state distributions of Fe for six days in 2006 with a one-hour time resolution\markmeold{ and in the same format as in Figure~\ref{fig:CSTimeseries}, with an additional panel for the magnetic-field polarity.} {Here, red indicates inwards pointing polarity ($\otimes$) and green indicates outwards-pointing polarity ($\odot$).}} 
}                               %
\end{figure}

\begin{figure}[htb!]\centering
%\begin{minipage}{0.45\textwidth}
  \includegraphics[width=\columnwidth]{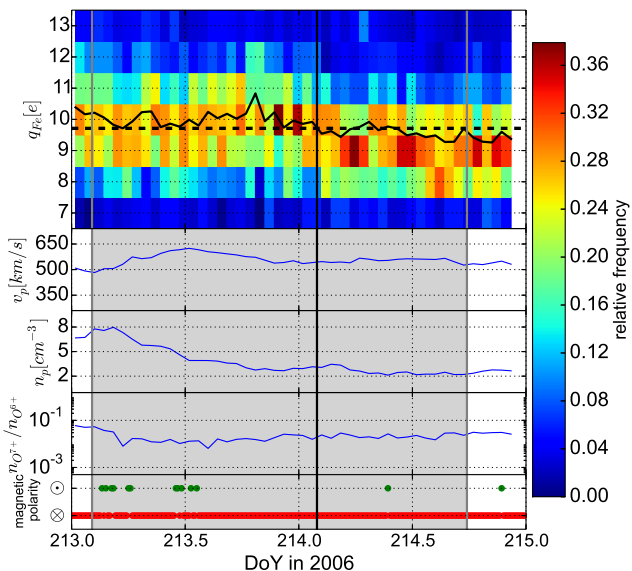}

\caption{\label{fig:coronal-hole-origin2} {Time series of the charge-state distributions of Fe for \markmeold{two} days in 2006, with a one-hour time resolution (first panel) \markmeold{in the same format as in Figure~\ref{fig:coronal-hole-origin}}.} 
}
\end{figure}

We \markme{now focus on the Fe charge-state distribution of individual coronal-hole-wind streams and relate the average Fe charge state of these} to the \markmeold{average} Fe charge-state of all slow solar wind streams\markme{ of the same year}.
\srem{\markmeold{From the {coronal-hole-wind} streams that were selected, \markme{three} time periods are \markme{discussed in detail} here: day of year (DoY) 2-14 in 2004 and DoY 158-162 and DoY 212-215 in 2006.}} \srem{In Figure~\ref{fig:coronal}, Figure~\ref{fig:CSTimeseries}, Figure~\ref{fig:coronal-hole-origin}, and Figure~\ref{fig:coronal-hole-origin2} some of the data excluded based on the selection criteria is plotted as well for the sake of providing context for the coronal-hole streams of interest.}

Figure~\ref{fig:coronal} summarizes the solar wind properties for 12 days in 2004. {From top to bottom, the panels in Figure~\ref{fig:coronal} show the collisional age $a_{\text{col}}$, the \markmeold{ratios $n_{O^{7+}}/n_{O^{6+}}$ and $n_{C^{6+}}/n_{C^{5+}}$}, the corresponding freeze-in temperatures $T_{f,{O^{7+}}/{O^{6+}}}$ and $T_{f,{C^{6+}}/{C^{5+}}}$, the magnetic field strength $B$ and angles $B_\phi$, $B_\theta$, and in the bottom panel the proton speed $v_p$ \markmeold{(left y-axis)} and proton density $n_p$ \markmeold{(right y-axis)}. }
A {coronal-hole stream} (DoY 3.5-6.5) is followed by an interface region \markmeold{with a higher-speed stream} and {an ICME beginning} on DoY 9. {The ICME period is marked with gray hatching.} From DoY 3.5 \markmeold{to DoY} 6.5, {$n_{O^{7+}}/n_{O^{6+}}$ and $n_{C^{6+}}/n_{C^{5+}}$} are low and, in particular, are {below their respective thresholds for coronal-hole wind}.  The collisional age $a_{\text{col}}$ is below {its threshold value of $0.1$ for coronal-hole wind} \markmeold{as well}. Thus, according to the criteria described in Section~\ref{sec:data}, % in \citet{zurbuchen2002solar,zhao2009global}, 
this suffices to categorize this stream as fast, that is, as {coronal-hole wind}. This is supported by the additional data products shown in Figure~\ref{fig:coronal}. Although there is some variability in the solar-wind proton speed $v_p$, the minimal value is still unlikely to be produced by slow solar wind.  %\sout{The alpha to proton ratio and the FIP ratio remain constant until $\sim$ DoY 6.5 where the onset of an interaction region is visible. The FIP (first ionization potential) ratio compares the ion to O abundance in the solar wind with the respective abundance observed in the photosphere (see \citet[]{vonSteiger2000}).
 {The freeze-in temperatures are derived under the assumption of an equilibrium state that allows us \markmeold{to} relate the observed abundance ratio of two adjacent ions \markmeold{to} the respective ionization and recombination rates: $n_{i}/n_{i+1}= R_{i+1}(T_f)/C_i(T_f)$. Since they depend on the density ratios $n_{O^{7+}}/n_{O^{6+}}$ and $n_{C^{6+}}/n_{C^{5+}}$ it is not surprising that the freeze-in temperatures are low during the coronal-hole-wind stream as well.}

Although still below the threshold \markmeold{for coronal-hole wind} defined above, the \markmeold{$n_{O^{7+}}/n_{O^{6+}}$} ratio and the collisional age are higher from DoY 3.5 to DoY 4.25 than in the following period. \markmeold{To prevent any interference by other processes, we focus on the part of the coronal-hole-wind stream from DoY 4.5-6.5 where both density ratios are safely below their respective categorization thresholds.}  Based on these considerations, the time period from DoY $4.5-6.5$ (which is indicated with the gray shaded area in Figure~\ref{fig:coronal}) contains only typical, quiet {coronal-hole wind}. 

\srem{Two days later, on \markmeold{DoY 9}, {an \markmeold{ICME} \markmeold{with a magnetic cloud}} follows.} {The ICME period exhibits a much larger variability in all data products; in particular, the {$n_{O^{7+}}/n_{O^{6+}}$}  and the collisional age are much higher. \markmeold{The magnetic-field angles show a smooth rotation indicating a magnetic cloud as part of the ICME.} Thus, this \markmeold{ICME} can easily be distinguished from the {coronal-hole-\markmeold{wind} stream}.}

\markmeold{Figure~\ref{fig:examples} focuses on the highlighted part of the coronal-hole-wind stream \markmeold{(DoY 4.5-6.5) from Figure~\ref{fig:coronal}.} For this time period of interest, } Figure~\ref{fig:examples} examines how the individual Fe charge states behave \markmeold{during DoY $4.5-5.6$ and DoY $5.6-6.5$. The motivation for this partition is detailed in Figure~\ref{fig:CSTimeseries}.} \srem{\markmeold{It} compares the Fe charge states of a period with Fe-cool (\markmeold{right, DoY $5.6-6.5$} in 2004) and Fe-hot (\markmeold{left, DoY $4.5-5.6$} in 2004) {coronal-hole wind}.} {The upper five panels \markme{of Figure~\ref{fig:examples}} show the density ratios} of C, O, and Fe ion pairs within {these} two \markmeold{parts of the} {coronal-hole-wind} stream\markmeold{ of interest,} and the {bottom} panel {provides examples} of the charge-state distributions of C, O, and Fe for two selected observations. The \markmeold{vertical} lines in the upper panels indicate the corresponding times used in the {bottom} panel. \markmeold{The horizontal lines in each panel give the mean density ratio in the left and right interval, respectively.}
All Fe ion density ratios {shown here} are higher \markmeold{in the first part} of the stream than in the \markmeold{second part}. \markmeold{Comparing the two charge-state distribution examples at the bottom}, \markmeold{in the example on the left}, not only is the maximum of the Fe charge-state distribution shifted from $Fe^{9+}$ to $Fe^{10+}$ \markmeold{ as compared to the second example,} but the complete distribution is shifted to higher charge states. {Thus, \markmeold{a} change in the mean charge-state cannot be explained by a single enhanced or depleted charge-state. Instead all considered charge states are affected.} 

%It is interesting to note, that in the Fe-hot {coronal-hole wind}, the ratios {$n_{O^{7+}}/n_{O^{6+}}$} and {$n_{C^{6+}}/n_{C^{5+}}$} are even lower than in the {Fe-cool coronal-hole wind} example.

\markmeold{As a continuous representation of the charge-state distributions in the bottom panel of Figure~\ref{fig:examples}, Figure~\ref{fig:CSTimeseries}} shows a time series of the \markmeold{Fe} charge-state distribution
 (considering only $Fe^{7+}$ to $Fe^{13+}$) \markmeold{in the first panel}. The three bottom panels
 provide the solar-wind proton speed $v_p$, proton density $n_p$, and
 the ratio \markmeold{$n_{O^{7+}}/n_{O^{6+}}$} as
 reference. \markmeold{During DoY 4.5 - 5.6} the charge-state
 distribution of Fe is shifted to higher charge states {more similar
   to those observed in slow solar wind. } \markmeold{A transition between
   Fe-hot and Fe-cool coronal-hole wind (marked with a vertical black
   line) occurs at DoY 5.6. }  \markmeold{After the transition, at
   $5.6-6.5$, lower Fe charge states are observed.} Both the
 characterization as Fe-cool or Fe-hot wind and the resulting
 transition point are defined on four-hour resolution data. \srem{Therefore,
 the 1-hour resolution data shown here, in
 Figure~\ref{fig:coronal-hole-origin} and in
 Figure~\ref{fig:coronal-hole-origin2} can cross the respective
 thresholds without contradicting the categorization.} \markmeold{The same transition divides the left and right parts of the five top panels in Figure~\ref{fig:examples}.} \markmeold{It is interesting to note that the Fe-hot part of the stream coincides with an average solar wind speed below 600 km/s, whereas the Fe-cool interval shows an average solar wind speed higher than 600 km/s. This hints at a potential stream boundary between two high-speed streams that coincides with the transition between Fe-hot and Fe-cool coronal-hole wind as a possible explanation for the observed transition.}

\begin{figure}[htb!]\centering

 \includegraphics[width=\columnwidth]{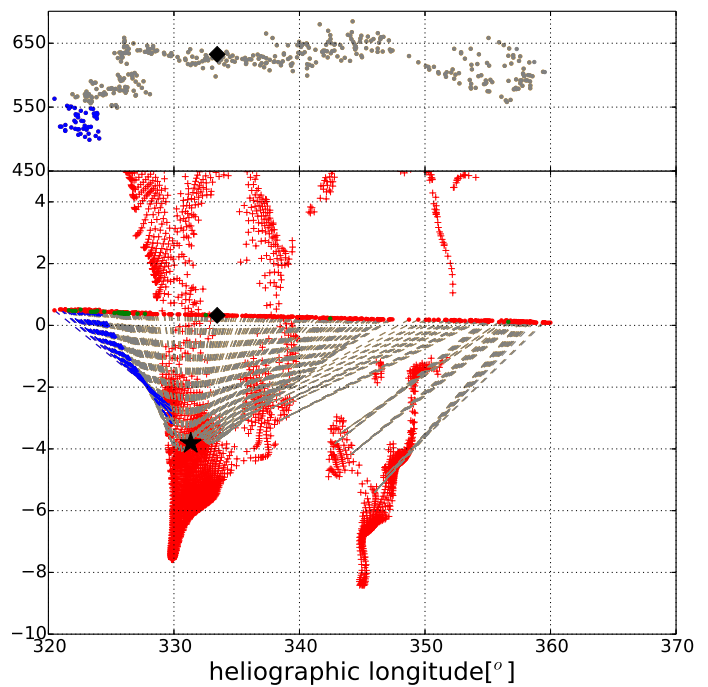}

\caption{\label{fig:CRmaps} {\markmeold{Section} of a heliographic map based on MDI magnetograms for Carrington rotation\markmeold{ 2044}.   \markmeold{The first part of the highlighted part of the stream in Figure~\ref{fig:coronal-hole-origin} is omitted because it is part of the previous Carrington rotation.} The \markmeold{top} panel shows the solar-wind proton speed ($v_p$) \markmeold{plotted against the mapped back} heliographic longitude on the source surface. Below, the dotted line shows the corresponding \markmeold{footpoints} of ACE for DoY \markmeold{158.4} - 162 on the source surface in heliographic coordinates.  In the the forthcoming A\&A online version, the color of each \markmeold{footpoint} on the source surface corresponds to the magnetic-field polarity observed at ACE. The polarity is plotted in 12-minute time resolution. These \markmeold{footpoints} are then traced down to the photosphere with a PFSS model. The dashed lines connect the positions of the ACE \markmeold{footpoints} on the source surface to the corresponding \markmeold{footpoints} on the photosphere. 
    \markmeold{\markmeold{Footpoints} of open field lines in the photosphere are indicated with $+$ symbols (dark-gray, i.e. red in the online version)\srem{ for Carrington rotation 2044}}.  The transition between the Fe-hot {coronal-hole wind} and the {Fe-cool coronal-hole wind} is marked with a black diamond on the source surface (and in the first panel) and with a black star on the photosphere. }
}
\end{figure}

\srem{\markmeold{For a solar wind stream in 2006 which exhibits a \markmeold{transition}
from {Fe-cool coronal-hole wind} to Fe-hot {coronal-hole wind}} Figure~\ref{fig:coronal-hole-origin} and Figure~\ref{fig:CRmaps}
{investigate} \markmeold{where the two} \markmeold{streams} \markmeold{originate} in \markmeold{the photosphere}.}

\markmeold{Figure \ref{fig:coronal-hole-origin} shows observations for a solar wind stream in 2006 which exhibits a \markmeold{transition} from Fe-cool CH wind to Fe-hot CH wind. } In the top panel of Figure \ref{fig:coronal-hole-origin}, a time series of the charge-state distribution is shown (in the same way as in
Figure~\ref{fig:CSTimeseries}). An increase of the average Fe charge state is visible at DoY {160.4} thus indicating a transition from {Fe-cool}
{coronal-hole wind} to Fe-hot {coronal-hole wind}. {The panels below
  \markmeold{show}} the solar-wind proton speed{, proton density, and the
  \markmeold{$n_{O^{7+}}/n_{O^{6+}}$} ratio measured at ACE.} \markmeold{To
  allow a direct comparison with the model polarity in
  Figure~\ref{fig:CRmaps}, t}he bottom panel \markmeold{gives}
additionally the magnetic-field polarity observed with ACE/MAG \markmeold{\citep{smith1998ace}} for the same
time period. In order to determine the in-situ magnetic-field polarity,
we first derive the nominal magnetic-field direction $B_{\phi}^{nom}$,
that is, the angle between the field line and the radial direction:
$B_{\phi}^{nom}=\arccos\left(\sqrt{\frac{1}{1+(\omega
    r\sin(\theta)/v_p)^2}}\right)$. Here, $\omega$ is the solar
angular velocity, $r$ is the Sun-spacecraft distance, $\theta$ is the
heliographic latitude and $v_p$ is the in-situ solar-wind proton
speed. Next, we \markmeold{subtract} the nominal magnetic field angle
$B_{\phi}^{nom}$ from the in-situ measured angle $B_{\phi}$. If the
absolute difference is greater than $90$ degrees, the magnetic-field polarity
is inwardly ($\otimes$, red) directed; otherwise it is outwardly ($\odot$,
green) directed.  \markmeold{A switch {from}
  \markmeold{outwardly pointing (i.e. $\odot$, green}) {polarity} to {inwardly pointing
    polarity (\markmeold{i.e.}  $\otimes$, red)} occurs at DoY $157.3$.}   From DoY $157.3$ onwards, the polarity remains {inwardly
  pointing ($\otimes$, red)} with some exceptions. We verified that
the exceptions are caused by kinks in the magnetic field which can be
seen by a reversal of the ion-proton differential streaming
\citep[]{berger2011systematic}. 
In particular, the
polarity does not change on DoY $160$ which includes the \markmeold{transition} from
{Fe-cool coronal-hole wind} to Fe-hot {coronal-hole wind}.  

%However, this
%  is not caused but the beginning of the transition to the following
%  slow solar wind stream but also coincides with kinks in the magnetic
%  field, i.e. both the coronal-hole-wind stream and the following slow
%  solar wind stream have the same polarity.}  

%Thus, it
%can be assumed that indeed both the normal {coronal-hole wind} and the
%following Fe-hot {coronal-hole wind} originate from the same coronal
%hole.

With a combination of ballistic back-mapping and a potential field
source surface (PFSS,
\citet[]{schatten-etal-1969,altschuler-and-newkirk-1969}) model, the
photospheric source region of the ACE observations from DoY
{$158-163$} can be estimated. Based on the in-situ solar wind speed,
the solar rotation, and the heliographic coordinates of ACE, the
position of ACE in heliographic coordinates is mapped back to the
source surface. Here, a simple PFSS model takes over and allows us to
track the field lines down to the photosphere. \markmeold{A uniform
  grid with 1$^{\circ}$ resolution is assumed at the source surface.}
\citet{peleikis2015sw14} gives a more detailed description of the
method applied here. The accuracy of this approach is limited by the
varying age of different parts of the underlying magnetograms that are
composites of images from 27 days. \markmeold{In the following,
\markmeold{this}  back-mapping is used to \markmeold{test} whether the observed coronal-hole wind can be associated with an open field line region in the
  photosphere and to investigate what kind of coronal structures are
  related to the transitions between Fe-cool and Fe-hot coronal-hole
  wind.} \markmeold{\markmeold{Although} only parts of the
  resulting heliographic maps are shown here, \markmeold{we examined
    the complete Carrington map in each case}.}

\markmeold{In Figure~\ref{fig:CRmaps}, each cross in the lower panel
  represents a \markmeold{footpoint} of a magnetic field line mapped back from
  the source surface down to the photosphere. Additionally for the corresponding Carrington
rotation\markmeold{ 2044}, the \markmeold{footpoints} 
of ACE are shown over this} Carrington map derived from a
PFSS model based on magnetrograms from the Michelson Doppler Imager
(MDI, \citet{scherrer1991solar}) on the Solar and Heliospheric Observatory (SOHO).  \markmeold{Figure~\ref{fig:CRmaps} shows only the area to
    which the observations in the stream of interest in
    Figure~\ref{fig:coronal-hole-origin} are mapped back by the
    PFSS model. This area is an extension of the northern polar coronal hole to
    equatorial regions} with \markmeold{- according to the PFSS model -}
  red, inwardly pointing polarity. The dashed
lines connecting the \markmeold{footpoints} on the source surface {and} the
photosphere corresponding to the \srem{fast} {coronal-hole-wind} stream from
DoY $158.12-161.59 $ are colored gray \markmeold{and the start of the subsequent declining phase of the stream (DoY $161.59-162$) is colored dark gray (blue in the on-line version).}
The border between the {Fe-cool coronal- hole wind} and the Fe-hot {coronal-hole wind} is indicated with a
black diamond on the source surface and with a black star on the
photosphere. 
\markmeold{The beginning of the period of interest belongs to the preceding Carrington rotation. Thus, for mapping the corresponding footpoints a different Carrington map would need to be considered. Since the transition itself occurs later in the stream and for the sake of clarity this first part of the stream is therefore omitted in Figure~\ref{fig:CRmaps}.}
{The \markmeold{remaining part of the} coronal-hole-wind stream of interest,
  \markmeold{including} the transition between the Fe-cool and the Fe-hot
  stream, and the following stream are mapped to a \markmeold{small
    scale region with open field lines} in the equatorial region which
  is\markmeold{, as we verified on the complete Carrington map,}
  connected to the northern polar coronal hole.}  {For the \markme{whole} time
  period and in particular, around the transition between the Fe-cool
  and Fe-hot stream, the in-situ observed polarities match (inwards
  pointing with the exception of the aforementioned kinks) the  polarities predicted by the
  PFSS model. However, due to
  the limitations on the accuracy of the back-mapping caused by the
  requirement to derive the magnetic field lines from a complete
  Carrington rotation, the possibility that the transition
  occurs at \markmeold{an} edge of this \markmeold{open field line region} cannot be ruled out.}

\begin{figure}[htb!]\centering

 \psfrag{sw}[][][1][0]{p}
 \includegraphics[width=\columnwidth]{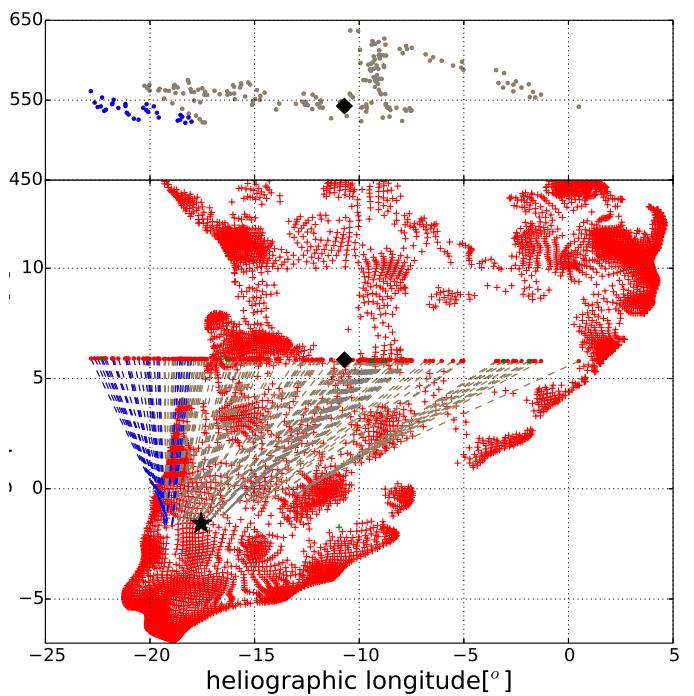}

\caption{\label{fig:CRmaps2} {A \markmeold{section} of a heliographic map based on an MDI magnetogram for Carrington rotation 2046. \markmeold{The format is the same as in Figure~\ref{fig:CRmaps}.}
  }
}
\end{figure} 

{A second transition, DoY 213-215 in 2006, in this case from an Fe-hot to an Fe-cool stream\markmeold{,} is shown  \markmeold{as a charge-state distribution time series in} Figure~\ref{fig:coronal-hole-origin2}. \markme{The Fe-hot part of this stream includes the beginning of the trailing edge of the high-speed stream as defined in \citet[]{borovsky2016trailing}.  The transition from Fe-cool to Fe-hot wind, however, occurs 12 hours earlier and is thus unlikely to be affected by the trailing edge.} %% and \ref{fig:CRmaps2}
\markmeold{It is interesting to note that the Fe-hot part of the stream of interest includes faster solar wind than the Fe-cool part of the stream. \markmeold{However, the change in solar wind speed occurs $>5$ hours earlier than the transition from Fe-hot to Fe-cool wind}. \markmeold{Since the transition boundary is defined on four-hour averages, shifting the temporal bins cannot make the two changes coincide exactly but they might be closer, as it appears here.}}
\markmeold{As \markmeold{shown} in Figure~\ref{fig:CRmaps2}, \markmeold{which provides a part of the photospheric map for the relevant Carrington rotation\srem{, in this case Carrington rotation} 2046,} the transition between the Fe-hot and Fe-cool coronal-hole-wind streams is situated within a larger region \markmeold{(compared to the case in Figure~\ref{fig:CRmaps})} of open field lines\markmeold{. Inspection of the heliographic map for the complete Carrington rotation (not shown here) indicates that this region was not connected to a polar coronal-hole.} In the following, we refer to such a region of open field lines\markmeold{ which is not connected to a polar coronal-hole }as isolated.} Within the accuracy of the back-mapping approach, the transition \markmeold{lies within a region of open field lines. However, within this region, the field line density is not uniform; for example west of the transition (to the right of the star in Figure~\ref{fig:CRmaps2}) the field line density is decreasing at the photosphere. This fine-structure could be related to the transition.} The in-situ magnetic polarity again matches the polarity predicted by the PFSS model with a few exceptions that can be explained as kinks in the magnetic field as well.} 
\markmeold{For both transition examples in Figures~\ref{fig:CRmaps} and~\ref{fig:CRmaps2}, both streams,} in particular also the Fe-\markmeold{hot} stream, originate in the equatorial region.

%\begin{table*}[htb!]\scriptsize\centering %for refereeversion
\begin{table*}[htb!]\small\centering
\caption{\label{tab:transitions} {Eight transitions between Fe-cool and Fe-hot coronal-hole-wind streams in 2006.}}
%\begin{tabularx}{\textwidth}{rlrccrrcrX}\hline\hline %for refereeversion
\begin{tabularx}{0.85\textwidth}{rlrccrrcrX}\hline\hline
  \markme{Stream of interest} & CR& Transition \markme{time} & \multicolumn{3}{c}{Latitude} & \multicolumn{3}{c}{\markmeold{Longitude}} & isolated or connected\\
  \markme{[DoY in 2006]} &  & \markme{[DoY in 2006]} & \multicolumn{3}{c}{\markme{[$^\circ$]}} & \multicolumn{3}{c}{\markme{[$^\circ$]}} & \\\hline
%  $26.9 - \phantom{0}28.4$ & 2039 & $-5.7 $&\markmeold{ to }&$ 8.1$ & 314.6 - 323.9& $\otimes$ isolated \\\hline
  $51.6 - \phantom{0}53.2$ & 2040 & \phantom{0}51.4 &$- 10.8 $&\markmeold{ to }&$ -5.7$ & 320.2 &to& 331.6& $\otimes$ isolated \\\hline
  $104.9 - 107.0$ & 2042  & 105.7 &$ -5.0$&\markmeold{ to }&  $-4.0$ &317.5 &to& 320.9 &$\otimes$ connected to NPCH \\\hline
  $131.5 - 134.4$ & 2043  & 133.4 & $ -7.7 $&\markmeold{ to }&$ -2.9$ & 326.7 &to& 343.9& $\otimes$ connected to NPCH \\\hline
   $*$ $158.1 - 161.6$& 2044  & 160.4 &$-5.4 $&\markmeold{ to }&$ -2.4$ & 329.8&to&348.5&$\otimes$ connected to NPCH\\\hline
  $186.2 - 188.1$ & 2045  & \markme{187.4} & $-4.9 $&\markmeold{ to }&$ -1.4$ & 331.1 &to& 343.5&$\otimes$ connected to NPCH \\\hline
   $*$ $213.1 - 214.5$ & 2046  & 214.1 & $-1.7 $&\markmeold{ to }&$ -2.9$ & 341.0 &to& 351.7&$\otimes$ isolated\\\hline
  $240.1 - 241.7$ & 2047  & 241.0 & $-0.1 $&\markmeold{ to }&$ 3.6$ &354.4 &to& 3.3&$\otimes$ isolated\\\hline
  $294.0 - 296.0$ & 2049  & 295.0 & $5.2 $&\markmeold{ to }&$ 5.7$ & 4.9 &to& 5.7&$\otimes$ isolated \\\hline
\end{tabularx}
\end{table*}

\begin{figure}[htb!]
  \includegraphics[width=\columnwidth]{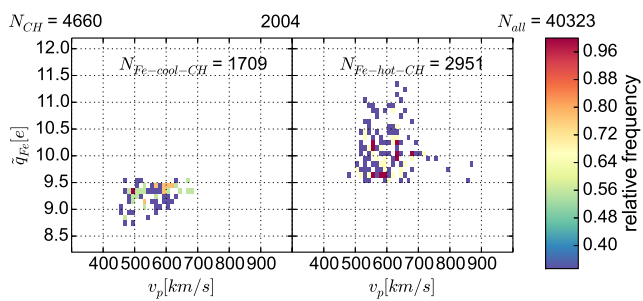}
\caption{\label{fig:overview2004} Average charge-state ($\tilde{q}_{Fe}$) versus solar wind proton speed for {Fe-cool} and Fe-hot coronal-hole type wind for 2004. The color gradient indicates the frequency of observing each charge-state-solar wind speed pair in 2004 and is normalized to the maximum in each panel. Each data point represents a four-hour average. {The inset in each panel gives the number of data points in 12-minute resolution that contributed to the averages in \markmeold{in this figure}. The number ($N_{all}$) of all 12-minute resolution data points in 2004 is given on the top right and the number ($N_{CH}$) of all data points categorized as pure coronal-hole wind, in the top left.}}
\end{figure}

\begin{figure}[htb!]
  \includegraphics[width=\columnwidth]{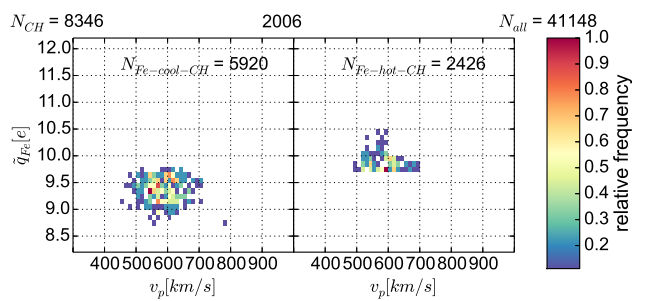}
\caption{\label{fig:overview2007} Average charge-state ($\tilde{q}_{Fe}$)   versus solar wind proton speed for {Fe-cool} and Fe-hot coronal-hole wind for 2006 \markmeold{in the same format as in Figure~\ref{fig:overview2004}}. 
}
\end{figure}

 Table~\ref{tab:transitions} summarizes the properties of eight transitions in 2006. The first column gives the start and end \markmeold{times of the part of the coronal-hole-wind stream containing the transition in DoY and} the corresponding Carrington rotation is noted in the second column. \markmeold{An $*$ in the first column indicates the transitions shown in Figures~\ref{fig:coronal-hole-origin}-\ref{fig:CRmaps2}.} \markmeold{The transition time is listed in the third column.} \markmeold{To indicate the latitudinal \markmeold{and longitudinal} position of the back-mapped foot points of each coronal-hole-wind stream, the highest and lowest back-mapped foot point is given in heliographic latitude in the \markmeold{fourth column, and the respective heliographic longitude in the fifth column.} T}he last column indicates whether the stream is mapped to \markmeold{an isolated open field line region} or whether the corresponding open field line region is connected to the northern polar coronal-hole (NPCH). All these coronal-hole-wind streams have inwardly pointing magnetic polarity as indicated by the symbol $\otimes$ in the last column. \markmeold{All eight transitions are mapped back to low latitudes and the longitudinal spread is at most $18^\circ$.}
As listed in Table~\ref{tab:transitions}, four out of eight transitions between Fe-cool and Fe-hot coronal-hole wind in 2006 were mapped back to isolated coronal holes in the equatorial region. The remaining three transitions were observed in the equatorial region as well, but the respective coronal holes were connected to the northern polar coronal-hole. With respect to \markmeold{the long-term behavior}, \citet[]{zhao2014polar} observed that wind from polar coronal holes is predominantly Fe-hot, while equatorial coronal-hole wind is predominantly Fe-cool. The case study shown here illustrates that,  \markmeold{independently of their connection to a polar coronal-hole}, Fe-hot coronal-hole-wind streams are hidden within the mainly Fe-cool equatorial coronal-hole wind.

\markmeold{A possible explanation for the change in the Fe charge states at transitions between Fe-cool and Fe-hot coronal-hole wind is that they coincide with stream interfaces between two distinct coronal-hole-wind streams. This is supported by the coinciding increase in solar wind speed and the decrease in proton density $n_p$ in the 2004 example in Figure~\ref{fig:CSTimeseries}. \markmeold{However, in Figures~\ref{fig:coronal-hole-origin} and~\ref{fig:coronal-hole-origin2}, no clear change in the proton density $n_p$ is visible at the transition. Furthermore, although the average solar wind speeds in the DoY 213-215 example in 2006  differ before and after the transition, the change in the solar wind speed does not coincide with the transition  but occurs more than five hours earlier. Within the accuracy of the back-mapping approach, the back-mapped positions of the transitions in Figures~\ref{fig:CRmaps} and~\ref{fig:CRmaps2} might be consistent with stream interfaces at the transition times but this is not conclusive.} Therefore, this explanation can neither be proved nor ruled out by the observations here.} %

\markmeold{To put the case studies presented above into some perspective, Figures~\ref{fig:overview2004} and~\ref{fig:overview2007} illustrate how frequently Fe-cool and Fe-hot coronal-hole-wind streams were \markmeold{observed in 2004 and 2006, respectively.}}
Figure~\ref{fig:overview2004} shows the frequency of the average charge-state versus the solar wind proton speed {for all Fe-cool coronal-hole wind in 2004 (left panel) and all Fe-hot coronal-hole wind in 2004 (right panel).} %
The inset in each panel indicates the number of {data points in 12-minute resolution of Fe-cool ($N_{Fe-cool-CH}$) and Fe-hot coronal-hole wind  ($N_{Fe-hot-CH}$) included in each panel}. \markmeold{For reference, the number $N_{all}$ of all 12-minute resolution data points in 2004 is given on the top right and the number $N_{CH}$ of all data points categorized as pure coronal-hole wind in the top left.} {In the same way, Figure~\ref{fig:overview2007} shows Fe-cool and Fe-hot coronal-hole wind for the year 2006.}  \markmeold{With respect to the solar wind speed, the distributions of both the Fe-cool and Fe-hot components of the coronal-hole wind overlap to a large extent in both years. In particular in 2004, the Fe-cool component contains slower solar wind than the Fe-hot component and the overlap is smaller than in 2006.} {The question \markmeold{under which conditions each type} of coronal-hole wind is prevalent \markmeold{in other years} is left for a later study. However, comparing the years 2004 (Figure~\ref{fig:overview2004}) and 2006 (Figure~\ref{fig:overview2007}) allows \markmeold{some} observations. Firstly, \markmeold{while in 2004, 1709 individual observations can be identified as Fe-cool coronal-hole wind and 2951 as Fe-hot coronal-hole wind, the 2006 data contains 5920 Fe-cool coronal-hole wind observations and 2426 Fe-hot coronal-hole wind observations.} This illustrates that in 2004 the Fe-hot wind is more frequent than the Fe-cool {coronal-hole wind}. {In 2006, however, the opposite is the case: Fe-cool coronal-hole wind is more frequent than Fe-hot coronal-hole wind.
\markmeold{ \markmeold{Secondly, not only does the frequency of each wind type change but the Fe-hot coronal-hole wind is less variable in its average Fe charge state and the overall} average Fe charge states are lower in 2006 than in 2004. (Although not shown here, this effect is not only visible for Fe but also for O.). } \markmeold{Both observations hint at a solar-cycle dependence  as observed for O and C in \citet[]{kasper2012evolution,schwadron2011coronal}.} Additionally, an overall drop of O and C charge states as observed in 2006\markmeold{, at the transition to the long solar minimum \markmeold{at the end of solar cycle 23}} \markmeold{(e.g. \citet[]{lepri2013solar,richardson2014identification}), \markmeold{is probably superimposed on the Fe charge-state distribution as well.}  This aspect requires further investigation.}

{The 2004 data \markmeold{in Figure~\ref{fig:overview2004}} exhibits an interesting feature with respect to \markmeold{a possible} solar-wind proton speed dependence of the average Fe charge state. The average Fe charge state of the Fe-cool component \markmeold{(left panel)} of the coronal-hole wind shows a \markmeold{possible} dependence on the solar-wind speed.  However, this is not visible in the Fe-hot component \markmeold{in the right panel} which instead manifests a larger variability in the average Fe charge state for each solar-wind speed. For the 2006 example, \markmeold{this feature is not visible}.

\section{Conclusions}
{Complementing the observations of \markmeold{the long-term behavior} of the Fe charge states in \citet[]{steiger2000composition,galvin2009solar,richardson2014identification,lepri2013solar, kasper2012evolution,schwadron2011coronal,zhao2014polar}, we present a case-study of individual} solar-wind streams that can be clearly identified as {coronal-hole wind,} {\markmeold{with either} high or low Fe charge states compared to the charge states in slow solar wind. Streams with either property occur in the same year and we \markmeold{also} \markmeold{observe} direct transitions between them.}  This indicates that the steady {coronal-hole wind} is less uniform {in terms of Fe charge states than with respect to O and C charge states.} In particular, solar-wind streams with high Fe charge states are Fe-hot and C/O-cool at the same time. Under the assumption that the charge-state distribution is frozen-in in the corona, a higher Fe charge state implies a higher freeze-in temperature {for the same cool O freeze-in temperature} and thus, a qualitatively different temperature profile in the corona. 

{The back-mapping of transitions between Fe-cool and Fe-hot coronal-hole-wind streams finds the origin of both streams in equatorial regions and close to each other. Thus, streams with consistently high \markmeold{or} low Fe charge states \markmeold{can} originate in the same region. A more systematic investigation of the origin of all individual Fe-hot and Fe-cool coronal-hole-wind streams is beyond the scope of this case-study. }
In addition, we have {seen} that in 2004 Fe-hot {coronal-hole wind} is {more frequent, while in 2006 Fe-cool coronal-hole wind is predominantly observed.} {A possible solar-cycle dependence of the frequency of Fe-hot and Fe-cool coronal-hole wind -- as it has been observed for O and C in \cite{schwadron2011coronal} and for Fe as well over solar cycle 23 in \cite{lepri2013solar} -- } {is one possibility to explain}  \markmeold{the changes in the frequency of Fe-hot and Fe-cool coronal-hole wind between 2004 and 2006.} 
{The details of a solar-cycle dependence}\markmeold{, the evolution of the respective coronal structures over consecutive Carrington rotations, and} the implications of these observations for the temperature profile in coronal holes {require further investigation}.

\begin{acknowledgements}
      Part of this work was supported by the 
      Deut\-sche For\-schungs\-ge\-mein\-schaft (DFG)\/ project
      number  Wi-2139/10-1\enspace. \markmeold{We very gratefully acknowledge the diligent work of the anonymous referee who provided very detailed and helpful suggestions and comments.}

\end{acknowledgements}

   \bibliographystyle{aa} % style aa.bst
   \bibliography{aa} % your references Yourfile.bib

\begin{thebibliography}{32}
\expandafter\ifx\csname natexlab\endcsname\relax\def\natexlab#1{#1}\fi

\bibitem[{{Altschuler} \& {Newkirk}(1969)}]{altschuler-and-newkirk-1969}
{Altschuler}, M.~D. \& {Newkirk}, G. 1969, Sol.~Phys., 9, 131

\bibitem[{Berger(2008)}]{berger2008velocity}
Berger, L. 2008, PhD thesis, Kiel, Christian-Albrechts-Universit{\"a}t, Diss.,
  2008

\bibitem[{Berger {et~al.}(2011)Berger, Wimmer-Schweingruber, \&
  Gloeckler}]{berger2011systematic}
Berger, L., Wimmer-Schweingruber, R., \& Gloeckler, G. 2011, Physical Review
  Letters, 106, 151103

\bibitem[{Borovsky \& Denton(2016)}]{borovsky2016trailing}
Borovsky, J.~E. \& Denton, M.~H. 2016, Journal of Geophysical Research: Space
  Physics

\bibitem[{Dere {et~al.}(1997)Dere, Landi, Mason, Fossi, \&
  Young}]{dere1997chianti}
Dere, K., Landi, E., Mason, H., Fossi, B.~M., \& Young, P. 1997, Astronomy and
  Astrophysics Supplement Series, 125, 149

\bibitem[{Galvin {et~al.}(2009)Galvin, Popecki, Simunac, Kistler, Ellis, Barry,
  Berger, Blush, Bochsler, Farrugia, {et~al.}}]{galvin2009solar}
Galvin, A., Popecki, M., Simunac, K., {et~al.} 2009, Ann. Geophys, 27, 3909

\bibitem[{Gloeckler {et~al.}(1998)Gloeckler, Cain, Ipavich, Tums, Bedini, Fisk,
  Zurbuchen, Bochsler, Fischer, Wimmer-Schweingruber,
  {et~al.}}]{gloeckler-etal-1998}
Gloeckler, G., Cain, J., Ipavich, F., {et~al.} 1998, in The Advanced
  Composition Explorer Mission (Springer), 497--539

\bibitem[{Gopalswamy {et~al.}(2009)Gopalswamy, Yashiro, Michalek, Stenborg,
  Vourlidas, Freeland, \& Howard}]{gopalswamy2009soho}
Gopalswamy, N., Yashiro, S., Michalek, G., {et~al.} 2009, Earth, Moon, and
  Planets, 104, 295

\bibitem[{Jian {et~al.}(2011)Jian, Russell, \& Luhmann}]{jian2011comparing}
Jian, L., Russell, C., \& Luhmann, J. 2011, Solar Physics, 274, 321

\bibitem[{Jian {et~al.}(2006)Jian, Russell, Luhmann, \&
  Skoug}]{jian2006properties}
Jian, L., Russell, C., Luhmann, J., \& Skoug, R. 2006, Solar Physics, 239, 393

\bibitem[{Kasper {et~al.}(2008)Kasper, Lazarus, \& Gary}]{kasper2008hot}
Kasper, J., Lazarus, A., \& Gary, S. 2008, Physical review letters, 101, 261103

\bibitem[{Kasper {et~al.}(2012)Kasper, Stevens, Korreck, Maruca, Kiefer,
  Schwadron, \& Lepri}]{kasper2012evolution}
Kasper, J., Stevens, M., Korreck, K., {et~al.} 2012, The Astrophysical Journal,
  745, 162

\bibitem[{Ko {et~al.}(1997)Ko, Fisk, Geiss, Gloeckler, \&
  Guhathakurta}]{ko1997empirical}
Ko, Y.-K., Fisk, L.~A., Geiss, J., Gloeckler, G., \& Guhathakurta, M. 1997,
  Solar Physics, 171, 345

\bibitem[{Landi {et~al.}(2013)Landi, Young, Dere, Del~Zanna, \&
  Mason}]{landi2013chianti}
Landi, E., Young, P., Dere, K., Del~Zanna, G., \& Mason, H. 2013, The
  Astrophysical Journal, 763, 86

\bibitem[{Lepri {et~al.}(2013)Lepri, Landi, \& Zurbuchen}]{lepri2013solar}
Lepri, S., Landi, E., \& Zurbuchen, T. 2013, The Astrophysical Journal, 768, 94

\bibitem[{Marsch(2006)}]{marsch2006kinetic}
Marsch, E. 2006, Living Reviews in Solar Physics, 3, 1

\bibitem[{McComas {et~al.}(1998)McComas, Bame, Barker, Feldman, Phillips,
  Riley, \& Griffee}]{mccomas1998solar}
McComas, D., Bame, S., Barker, P., {et~al.} 1998, in The Advanced Composition
  Explorer Mission (Springer), 563--612

\bibitem[{Peleikis {et~al.}(2015)Peleikis, Kruse, Berger, Drews, \&
  Wimmer-Schweingruber}]{peleikis2015sw14}
Peleikis, T., Kruse, M., Berger, L., Drews, C., \& Wimmer-Schweingruber, R.~F.
  2015, in Solar Wind 14 Proceedings

\bibitem[{Richardson(2014)}]{richardson2014identification}
Richardson, I. 2014, Solar Physics, 289, 3843

\bibitem[{Richardson \& Cane(2010)}]{richardson2010near}
Richardson, I. \& Cane, H. 2010, Solar Physics, 264, 189

\bibitem[{Schatten {et~al.}(1969)Schatten, Wilcox, \&
  Ness}]{schatten-etal-1969}
Schatten, K., Wilcox, J., \& Ness, N. 1969, Sol.~Phys., 6, 442

\bibitem[{Scherrer {et~al.}(1991)Scherrer, Hoeksema, \&
  Bush}]{scherrer1991solar}
Scherrer, P., Hoeksema, J., \& Bush, R. 1991, Advances in Space Research, 11,
  113

\bibitem[{Schwadron {et~al.}(2011)Schwadron, Smith, Spence, Kasper, Korreck,
  Stevens, Maruca, Kiefer, Lepri, \& McComas}]{schwadron2011coronal}
Schwadron, N., Smith, C., Spence, H.~E., {et~al.} 2011, The Astrophysical
  Journal, 739, 9

\bibitem[{Smith {et~al.}(1998)Smith, L’Heureux, Ness, Acu{\~n}a, Burlaga, \&
  Scheifele}]{smith1998ace}
Smith, C.~W., L’Heureux, J., Ness, N.~F., {et~al.} 1998, in The Advanced
  Composition Explorer Mission (Springer), 613--632

\bibitem[{Tu {et~al.}(2005)Tu, Zhou, Marsch, Xia, Zhao, Wang, \&
  Wilhelm}]{Tu2005}
Tu, C., Zhou, C., Marsch, E., {et~al.} 2005, Science, 308, 519

\bibitem[{von Steiger {et~al.}(2000)von Steiger, Schwadron, Fisk, Geiss,
  Gloeckler, Hefti, Wilken, Wimmer-Schweingruber, \&
  Zurbuchen}]{steiger2000composition}
von Steiger, R.~v., Schwadron, N., Fisk, L., {et~al.} 2000, {J}ournal of
  {G}eophysical {R}esearch: Space Physics (1978--2012), 105, 27217

\bibitem[{Wilhelm(2012)}]{wilhelm2012sumer}
Wilhelm, K. 2012, Space science reviews, 172, 57

\bibitem[{Yashiro {et~al.}(2004)Yashiro, Gopalswamy, Michalek, St~Cyr,
  Plunkett, Rich, \& Howard}]{yashiro2004catalog}
Yashiro, S., Gopalswamy, N., Michalek, G., {et~al.} 2004, Journal of
  Geophysical Research: Space Physics, 109

\bibitem[{Zhao \& Fisk(2010)}]{zhao2010comparison}
Zhao, L. \& Fisk, L. 2010, in SOHO-23: Understanding a Peculiar Solar Minimum,
  Vol. 428, 229

\bibitem[{Zhao \& Landi(2014)}]{zhao2014polar}
Zhao, L. \& Landi, E. 2014, The Astrophysical Journal, 781, 110

\bibitem[{{Zhao} {et~al.}(2009){Zhao}, {Zurbuchen}, \& {Fisk}}]{zhao2009global}
{Zhao}, L., {Zurbuchen}, T.~H., \& {Fisk}, L.~A. 2009, Geophysical Research
  Letters, 36, L14104

\bibitem[{Zurbuchen {et~al.}(2002)Zurbuchen, Fisk, Gloeckler, \& von
  Steiger}]{zurbuchen2002solar}
Zurbuchen, T., Fisk, L., Gloeckler, G., \& von Steiger, R. 2002, Geophysical
  {R}esearch {L}etters, 29, 66

\end{thebibliography}

\end{document}